\documentclass[fp]{jpsj3}
\usepackage{txfonts}

\title{Hysteretic transition of coarse-grained power grid model on small-world network}
\author{Eiichi SASAKI$^1$\thanks{esasaki@me.es.osaka-u.ac.jp},
	Masayuki OHZEKI$^2$ and	Yoshito OHTA$^{3}$}
\inst{$^1$ Graduate School of Engineering Science, Osaka University, Osaka 560-8531, Japan\\
$^2$Department of Systems Science, Graduate School of Informatics, Kyoto University, Kyoto 606-8501, Japan \\
$^3$Department of Applied Mathematics and Physics, Graduate School of Informatics, Kyoto University, Kyoto 606-8501, Japan, JST-CREST}

\abst{
 We study synchronous phenomena of a coarse-grained power grid model, the swing equation, 
 on small-world networks.
 We show that its steady state, which stands for the normal operation of the power systems, 
 can be realized even if the phases are disordered. 
 In addition, we prove the linear stability of steady state 
 with small-different phases between the adjacent oscillators.
 On the other hand, a trigger of instantaneous power failure, which is described 
 by the hysteretic transition, might disappear on an appropriate small-world network.
 This result suggests that the small-world connection would potentially prevent the massive blackouts.
}

\kword{Synchronization, Hysteresis, Complex network}

\begin{document}
\maketitle

\section{Introduction}

  The steady electric power supply and proper operation of power systems have been an important problem 
  to support our daily life and avoid massive blackouts$^{1)}$. 
  On the other hand, the fundamental understanding on the stable optimized operation of the power system 
  has not been established sufficiently because the dynamics of the power system includes too complicated phenomena.

  From a point of view of the complex network analysis, the power grid is considered as a typical example of 
  the large artificial networks, and its statistical properties of transmission networks have been investigated.
  Pagani and Aiello$^{2)}$ 
  reviewed a lot of researches applying the complex network analysis 
  to various real power grids.
  They showed that the real power grids consists of $\sim 10^4$ nodes, which connect 
  with edges of transmission lines, and their mean degrees, the mean numbers of edges of 
  transmission lines per a node, 
  are in the range of $2\sim4$, respectively.
  If the mean distance is small and the cluster coefficient is large the network becomes a small world, 
  which has been considered 
  as a distinctive character of natural networks$^{3)}$. 
  However, in the case of the power grids, the mean distance is small but the cluster coefficient variously changes.
  In other words, it has not been clear whether there are benefits of small world for the power grid network.

  In the present study we discuss the stability and transition of power systems 
  on the small-world network generated on the basis of  the Watts-Strogatz model$^{4)}$. 
  The model creates the small world by the random reconnection of the regular loop.
  We deal with the swing equation, which is a coarse-grained model on the phases of the alternating currents 
  in the power system$^{5,6)}$, 
  on the Watts-Strogatz model.
  It should be noted that the swing equation without  the inertia term corresponds 
  to the Kuramoto model$^{7)}$, 
  which is a solvable oscillator model of synchronous phenomena.
  The swing equation has thus been attracting many researchers' interest not only as a model 
  of power grids but also as an extended problem of Kuramoto oscillators.
  Indeed the swing equation has been analyzed by Tanaka et al.$^{8)}$ 
  in particular on its phase transition on the complete graph for various torque distributions.
  They found that the oscillators have the first-order hysteretic phase transition, the process 
  of which depends on the type of the torque distributions and the mean-field approximation gives 
  good predictions on the phase transition.
  The synchronous conditions of the swing oscillators was studied by D\"orfler and Bullo$^{9)}$ 
  through the singular perturbation from the Kuramoto oscillators.
  They proved that the oscillators converge to the phase-lock solution of Kuramoto oscillators 
  when the mass of swing oscillators is sufficiently small.
  The simplicity of the swing equation drives other researches such as 
  the distributed power generating system$^{10)}$ 
  and the stability of steady states on the real power networks$^{11)}$. 

  In order to compare the swing equation to the Kuramoto oscillators we briefly review the studies 
  on the Kuramoto oscillators. 
  The phase transition of the Kuramoto oscillators on the Watts-Strogatz network was 
  studied by Hong et al.$^{12)}$. 
  They reported that the oscillators appear the second-order phase transition with the frequency 
  synchronization similar to that on the complete network.
  On the complete graph, the process of phase transition depends on the type of 
  frequency distribution$^{7)}$, 
  which differs from the case of random graph.
  Um et al.$^{13)}$ 
  studied the phase transition of Kuramoto oscillators on 
  the random graph varying the type of the frequency distribution.
  They found that since the random connections violate the transition characters 
  the oscillators have the common second-order phase transition regardless of the frequency distributions.

  In this paper, we discuss the stability and phase transition of the swing equation 
  on the Watts-Strogatz model to study properties of the power systems on 
  the small-world network.
  Although the analysis on the small world following Watts-Strogatz model is not enough 
  to directly understand the real power grid, we focus on the stability and 
  transition of swing equation on the Watts-Strogatz model as a simplified model of the real power grid.

  The remaining of the paper is organized as follows. 
  The second section fixes the notation and definition of the swing equation and the Watts-Strogatz model. 
  In the third section we prove the linear stability of the steady state with the small-different 
  phases between adjacent oscillators.
  This stability allows a different type of the steady state from that with the well-known synchronization.
  In the forth section we demonstrate several numerical computations and report the following results.
  In the case of the random graph, the oscillators have the common first-order hysteretic phase transition 
  between the ordinary steady state with synchronization and the unsteady state regardless of 
  the type of the torque distribution.
  In the case of small-world network, on the other hand, the oscillators become the steady states 
  with small-different phases without synchronization, and the hysteretic transition seems to disappear 
  on an appropriate small-world network.
  This fact implies that on the small-world network we might be capable to operate the power system 
  without fear of the blackout. 
  The fifth section is devoted to conclusion and discussion.
  
\section{Problem setting}

  We deal with the dimensionless swing equation on a small-world network.
  The phase of $i$th-generator/consumer, $\theta_i\in[0,2\pi)$, is governed by
  \begin{align}
   m\ddot{\theta}_i+\dot{\theta}_i
   =
   \omega_i+K\sum_j^N A_{ij} \sin(\theta_j-\theta_i),\label{swing}
  \end{align}
  where $\ddot{\theta}_i=d^2\theta_i/dt^2$ and $\dot{\theta}_i=d\theta_i/dt$ are the time derivatives.
  Here $t$ denotes the time, and $N$ represents 
  the total number of generators and consumers$^{5,6)}$. 
  The steady state, $\dot{\theta}_i=0(i=1,\cdots N)$, represents the normal operation 
  in the context of the power systems.
  On the other hand the unsteady state, $\dot{\theta}_i\neq0$, signifies the blackout. 
  In addition $m$ and $K$ are the mass and the coupling constant, respectively.
  The torque $\omega_i$ follows a distribution $g(\omega)$ with a vanishing mean.
  Then  $\omega_i>0$ indicates the generator providing energy to the power systems, 
  and $\omega_i<0$ stands for the consumer expending the energy transferred from the generators.
  It is noted that the zero mean of the torque distribution
  is necessary for the existence of steady states.
  The second term on the right-hand side of the equation \eqref{swing} 
  indicates the energy transfer between each node, generator and consumer, 
  and $A_{ij}$ is the $(i,j)$ element of the adjacency matrix $A$.
  If the node $i$ and $j$ are connected $A_{ij}=1$ and otherwise $A_{ij}=0$.
  The network is assumed to be undirected with neither loop nor multiedge.
  Then, $A$ is a symmetric matrix without any diagonal elements, $A_{ij}=A_{ji}$ and $A_{ii}=0$.
  The degree of $i$th node is given by $k_i=\sum_jA_{ij}$, 
  and the mean degree is obtained by $\left\langle k\right\rangle=\sum_ik_i/N$.
  We note that, from a point of view of the phase reduction theory, the inertial term, 
  $m\ddot{\theta}_i$, is {\it ad hoc} but it is necessary for describing 
  abrupt phenomena on the power grids 
  such as the instantaneous power failure$^{15)}$.

  In order to investigate the behavior of the phase following the swing equation 
  on a typical small-world network, we employ the Watts-Strogatz model$^{4)}$. 
  The model can generate various small-world networks by modifying the regular network 
  with the reconnection probability $P$.
  In the case of $P=0$ the network is regular, and the mean distance and cluster coefficient are large.
  On the other hand, in the case of $P=1$, the network changes into 
  the random graph with the short mean distance and small cluster coefficient.
  In the range of $0<P<1$ the network becomes small world; 
  the small mean distance and large cluster coefficient.
  Notice that, on the real power grids, the mean distance is small 
  but the cluster coefficient can variously change$^{2)}$. 
  
\section{Linear stability of steady states with small-different phases}

  In this section, we first prove the linear stability of a steady state 
  with small-different phases between adjacent oscillators on an undirected network,
  \begin{align*}
   0=\omega_i+&K\sum_{j=1}^NA_{ij}\sin(\theta_j^0-\theta_i^0)
   \;\;\;
   i=1,\cdots,N,\\
   &|\theta_i^0-\theta_j^0|<\frac{\pi}{2}
   \;\;\;
   {\rm for}\;\;\;\{(i,j)|A_{ij}=1\}\nonumber.
  \end{align*}
  Substituting $\theta_i=\theta_i^0+\hat{\theta}_i\exp(\sigma t)$ to \eqref{swing} and 
  neglecting the second order term of $\hat{\theta}_i$, we obtain a linearized equation of $\theta_i^0$,
  \begin{align*}
   \sigma^2m\theta_i+\sigma\theta_i=
   K\sum_{j=1}^N A_{ij}\cos(\theta_j^0-\theta_i^0)(\theta_j-\theta_i),
  \end{align*}
  where $\hat{\;\cdot\;}$ is omitted for simplicity.
  Taking a product with $\theta_i$ over $i$ leads to the energy equation of infinitesimal disturbance,
  \begin{align}
   \sigma^2
   \left(
   \frac{m}{N}\sum_{i=1}^N\theta_i^2
   \right)
   +
   \sigma
   \left(
   \frac{1}{N}\sum_{i=1}^N\theta_i^2
   \right)
   +
   \frac{K}{N}
   \sum_{i=1}^N\sum_{j=i+1}^N
   A_{ij}\cos(\theta_j^0-\theta_i^0)(\theta_j-\theta_i)^2=0\label{energy}.
  \end{align}
  Here we use the fact $A_{ij}=A_{ji}$.
  Since the equation \eqref{energy} is an algebraic equation of the second degree 
  with respect to $\sigma$, we obtain the roots
  \begin{align*}
   \sigma_{\pm}
   =
   \frac{1}{2m}
   \left(
   -1\pm\sqrt{1-4m^2K\frac{P}{V}}
   \right),
  \end{align*}
  where $V$ and $P$ are the kinetic energy and potential,
  \begin{align*}
   V
   =
   \frac{m}{N}\sum_{i=1}^N\theta_i^2>0,
   \;\;\;
   P
   =
   \frac{1}{N}\sum_{i=1}^N\sum_{j=1}^N
   A_{ij}\cos(\theta_j^0-\theta_i^0)(\theta_j-\theta_i)^2>0.
  \end{align*}
  Therefore ${\rm Re}[\sigma_{\pm}]<0$.
  We conclude that a steady state with small-different phases 
  between adjacent oscillators on an undirected network is linearly stable.
  It should be emphasized that the steady state does not necessarily imply the phase synchronization,
  although the steady state with small-different phases includes the ordinary phase lock state.
  D\"orfler and Bullo$^{9)}$ 
  analytically studied the phase-lock state 
  of the swing equation using the singular perturbation from the Kuramoto oscillator ($m=0$), 
  and they proved that, if $m$ is sufficiently small and 
  $\displaystyle NK>\max_{1\leq i\leq j \leq N}(\omega_i-\omega_j)$, 
  the oscillator asymptotically converges to a phase-lock solution of Kuramoto oscillator 
  with the order of $O(m)$.
  Notice that our result is free from the limitation on $m$.

  The remaining problem is where such a steady state exists. 
  In the next section we show several numerical results to demonstrate 
  that this different type of the steady state with small-different phases, 
  the linear stability of which is here proven, appears as the reconnection probability gradually decreases.

\section{Phase transition on a small-world network}

  In the present section we perform numerical integration of the swing equation 
  and show the phase transition of the swing oscillators on the Watts-Strogatz network.
  We then take three types of the torque distributions; 
  the uniform distribution, Gaussian and Bimodal Gaussian, 
  \begin{align*}
   g_U(\omega)&=
   \left\{
   \begin{array}{ll}
    1/2 & |\omega|\leq1\\
    0   & |\omega|>1
   \end{array}
   \right.,\\
   g_G(\omega)&=\frac{1}{\sqrt{2\pi}}\exp\left(-\frac{\omega^2}{2}\right),\\
   g_{BG}(\omega)&=\frac{1}{2}(g_G(\omega-1)+g_G(\omega+1)).
  \end{align*}
  The numerical integration is performed by the second-order Adams-Bathforth method.
  We take the time step $\delta t=0.05$ and increase 
  the end time of integration up to $T_f=10^4$ to reach the steady state. 
  The initial condition of phases and angular velocities is generated 
  from uniform distributions in the range of $[-1/2,1/2]$.
  In order to evaluate the phase transition of the swing oscillators 
  we estimate the order parameter, $R$, which characterizes the phase-lock state.
  \begin{align*}
   R=
   \frac{1}{T_w}
   \int_{T_s}^{T_f}
   \left|\frac{1}{N}\sum_{i=1}^N\exp(i\theta_i(t))\right|dt,
  \end{align*}
  where $T_s=200$ and $T_w=T_f-T_s$.
  When the oscillators are in a phase-locked state, $R\simeq1$.
  On the other hand, in the disordered state,  $R\simeq O(1/N)$.
  In order to check the steadiness of the state we compute the squared angular velocity,
  \begin{align*}
   \left \langle \dot{\theta}^2 \right \rangle
    =
    \frac{1}{T_w}
    \int_{T_s}^{T_f}
    \left(
     \frac{1}{N}\sum_{i=1}^N \dot{\theta}_i^2 
    \right)
    dt.
  \end{align*}
  When the squared angular velocity vanishes, the steady state, 
  the normal operation of the power systems, is realized.
  On the other hand the unsteady state with nonzero squared angular velocity signifies the blackout. 
  In the case of $K=0$, the phase states are given by $\theta_i(t)=\omega_i t+C_{i1}\exp(-t/m)+C_{i2}$
  where $C_{i1}$ and $C_{i2}$ are constants depending on the initial condition.
  If $K$ is small the squared angular velocity is about the variance of the torque distribution, 
  $\int \omega^2g(\omega)d\omega$.
  The number of oscillators is up to $N=10^4$.
  We take the ensemble average over 100 samples of the randomly generated networks and torques.

\begin{figure}
   \centering
   \includegraphics[width=150mm]{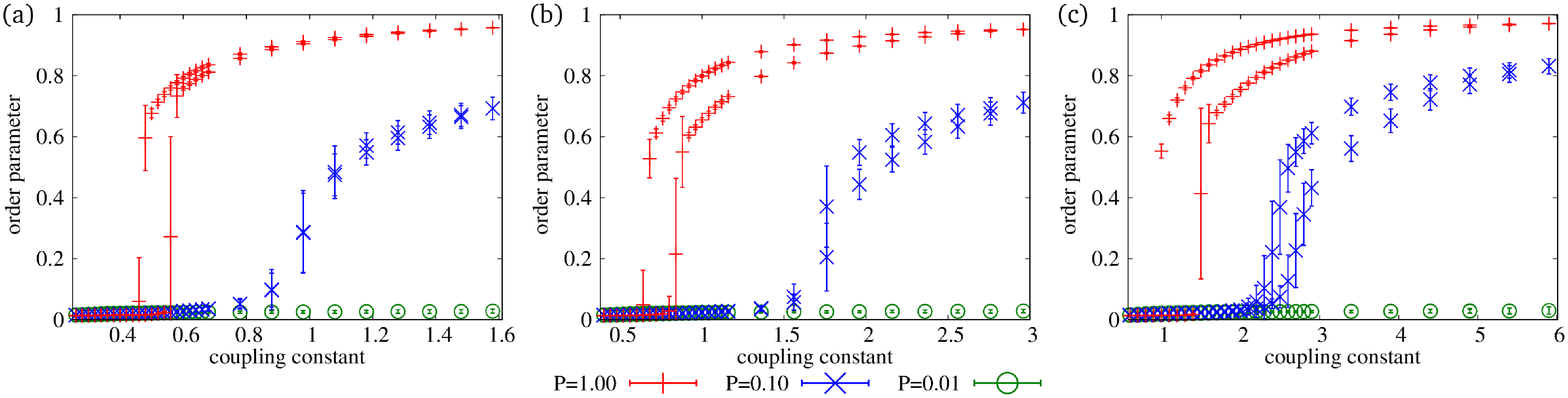}
   \caption{(Color online) Order parameter ($N=10^4,T_f=10^4$):
   (a) uniform,
   (b) Gaussian and (c) bimodal Gaussian torque distributions.
   The horizontal and vertical axes indicate the coupling constant 
   and the order parameter $R$, respectively.
   The red crosses, blue saltires and green circles indicate
   the cases of $P=1.00,0.10$ and $0.01$, respectively.
   The error bars are the standard deviation of realizations.
   }
   \label{fig:order}
   \centering
   \includegraphics[width=150mm]{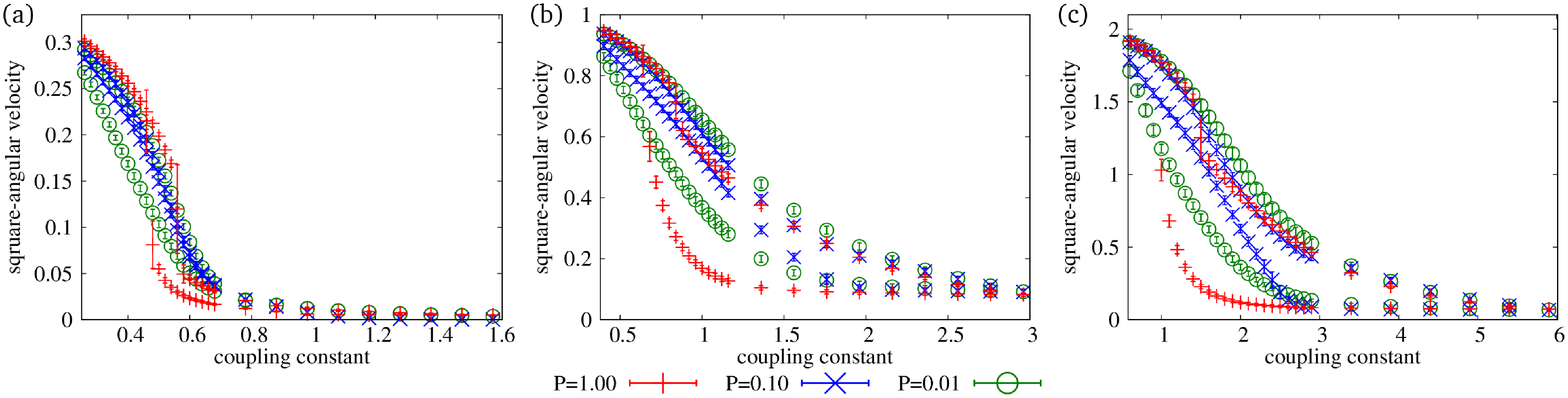}
   \caption{(Color online) Same as fig.\ref{fig:order} but the squared angular velocity ($N=10^4,T_f=10^4$).
   }
   \label{fig:mean_vel}
  \end{figure}

  Figure \ref{fig:order} shows the order parameter for each torque distribution,
  while we plot the squared angular velocity in Fig. \ref{fig:mean_vel} in order to check steady states.
  We here show the representative results of the case $m=5$ and $\left\langle k\right\rangle=4$.
  We remark that we have confirmed that the results do not qualitatively change 
  with varying $m$ and $\left\langle k\right\rangle$.
  In addition the results do not qualitatively change by varying the end time and the system size.

  In the case of the random graph, $P=1.00$, as the coupling constant increases, the order parameter 
  increases discontinuously and the oscillators become the phase-lock state with $R\sim 1$.
  On the other hand, when we gradually decrease the coupling constant, the order parameter falls 
  discontinuously and the disordered state appears.
  As the order parameter changes discontinuously between $O(1/N)$ and $O(1)$, 
  the squared angular velocity also changes discontinuously between $\int \omega^2g(\omega)d\omega$ 
  and $O(0)$ as in Fig. \ref{fig:mean_vel}.
  The coupling constant of transition from the disordered unsteady state to 
  phase-lock steady state is larger than its inverse.
  Thus, the oscillators have the first-order hysteretic phase transition in the case of the random graph.
  Even if the torque distribution is different the transition process seems quantitatively unchanged.

  On the other hand, on the small-world networks, which is generated by gradually decreasing 
  the reconnection probability, the transition process becomes smooth compared to the case on the random graph. 
  In the region with the high coupling constants the squared angular velocity becomes 
  as small as that of the random graph.
  In particular let us focus on the case with $P=0.01$.
  The order parameter hardly varies
  from $O(1/N)$,  while, as the coupling constant changes,
  the squared angular velocity changes between $\int \omega^2g(\omega)d\omega$ 
  and $O(0)$ through the different paths.
  Notice that the squared angular velocity exhibits the smooth hysteretic transition.
  This fact ensures that the steady state can be realized although the order parameter is a small value.
  As proven in the previous section, the steady state with small-different phases is linearly stable.
  In other words, the small-world network admits the steady state different from the ordinary phase-lock state.

  In addition, the transition processes of phases and squared angular velocity are clearly different 
  from that of the random graph.
  We have to emphasize that the transition point of the squared angular velocity is different from 
  that of the order parameter.
  Although the order parameter is often investigated to identify the nontrivial phase transition, 
  the steadiness is also important to characterize the behavior of the oscillators.  
  In the context of the power system, the steadiness should be rather highlighted since 
  the steady state represents the normal operation.
  In this sense, the hysteretic transition should be avoided in order to keep 
  the normal operation on the small-world network.
  Let us closely investigate the existence of the hysteretic transition in the squared angular velocity below.

\begin{figure}
   \centering
   \includegraphics[width=150mm]{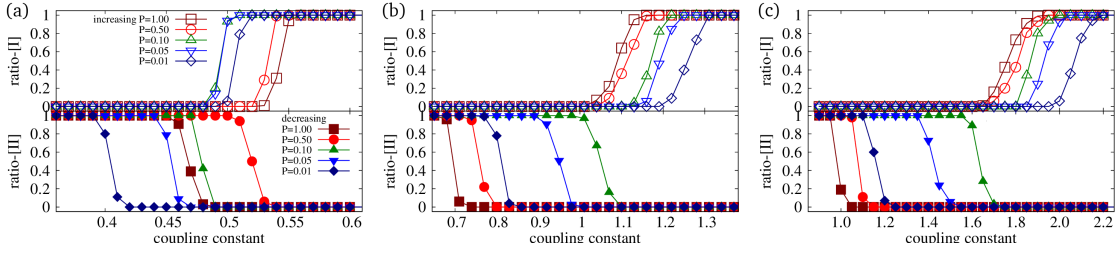}
   \caption{(Color online) Rate of transition($N=10^4,T_f=10^4$):
   (a) uniform,
   (b) Gaussian and (c) bimodal Gaussian torque distributions.
   The top and bottom columns show increasing and decreasing processes, respectively.
   The maroon squares, red circles, green triangles, blue-reverse triangles 
   and navy rhomboids indicate the cases of $P=1.00,0.50,0.10,0.05$ and $0.01$, 
   respectively.
   The horizontal and vertical axes indicate the coupling constant 
   and the ratio of transition, respectively.
   }
   \label{fig:smallworld_rate}
  \end{figure}

  We compute a rate of transition samples in order to investigate smooth hysteretic process of 
  the squared angular velocity in detail.
  We set the threshold which judges a sample in the steady state or not at $\int \omega^2g(\omega)d\omega/2$, 
  which is the half of the squared angular velocity at $K=0$.
  As the coupling constant increases/decreases we count the number of samples 
  of which the squared angular velocity is more/less than this threshold.
  The rate of transition is given by dividing this number of counted samples by the number of all samples.
  Figure \ref{fig:smallworld_rate} shows the rate of transition for each torque distribution.
  The rate of transition is a near step function, and the critical points of 
  transition could be distinguished clearly.
  We find that as the reconnection probability decreases the difference 
  between two critical coupling constants decreases until $P\simeq0.1$, 
  and as the reconnection probability further decreases this difference increases monotonically.
  It should be noticed that on Fig. \ref{fig:mean_vel} in the case of $P=0.1$ 
  the difference of the hysteretic paths is smaller than that of the cases of $P=1.0$ and $0.01$.
  This tendency is consistent with the results of the rate of transition.
  These results imply that for any torque distribution the hysteretic process 
  of the squared angular velocity would change and possibly disappear around $P\simeq0.1$.
  It is noted that we have obtained this result using $10^4$ samples and $N=500$.
  We restrict ourselves to just saying possibly that the hysteretic 
  transition disappears on some small-world network.
   
  In addition, we study the steady state at the extremely high coupling constant
  and find that, as the reconnection probability decreases,
  the order parameter decays but its degrade is not significant (not shown).
  This result suggests that the phase-lock state is realized regardless of the reconnection probability 
  as long as the coupling constant is sufficiently large value.
  We also study the correlation of the phases of steady oscillators,
  and we find that the oscillators distant about the mean distance of network are correlated (not shown).
  It is noted that, as the reconnection probability decreases, the mean distance increases.
  This result implies that the correlation between distant oscillators would lead to 
  the variety of phases of steady state. 
  Especially, the steady state without the phase synchronization
  would arise from the slightly longer distant correlation rather than that of the random graph, 
  such as the case of $P=0.01$ shown in Figs. \ref{fig:order} and \ref{fig:mean_vel}.

\section{Conclusion and discussion}

  We study the phase transition of swing equation on a small-world network.
  We prove the linear stability of the steady states 
  with small-different phases between adjacent oscillators.
  This result permits the steady state without the phase lock.

  Next, we study the phase transition on the Watts-Strogatz network for various torque distributions.
  In the case of the random graph,
  we find that the oscillators have the 1st-order hysteretic phase transition.
  The transition process seems to be quantitatively unchanged by the type of torque distributions.
  On the other hand Um et al.$^{13)}$ 
  found that the Kuramoto oscillators on the random graph have 
  the common 2nd-order phase transition regardless of the type of the frequency distributions.
  This result suggests that
  on the random graph the uniform phase transition independent of the distribution 
  is a common property between the swing and Kuramoto oscillators.
  In addition, we have employed 
  the mean field approximation$^{14)}$ 
  but the approximation does not work well.
  Um et al.$^{13)}$ 
  reported that the mean field approximation for 
  the Kuramoto oscillators on the random graph is not consist with the numerical results.
  Therefore this result implies that the earlier mean field method could not deal with
  the random distributed interactions.

  In the case of the small-world  network, 
  we find that as the reconnection probability decreases
  the transition process of the squared angular velocity 
  becomes discontinuous to smooth.
  In spite of incoherent phases the oscillators become the steady state
  as the coupling constant increases.
  Moreover, at the marginal reconnection probability, 
  the hysteretic transition of the squared angular velocity seems to disappear.
  Hong et al.$^{12)}$ 
  studied the Kuramoto oscillators on the Watts-Strogatz network 
  and found that the 2nd-order phase transition of oscillators accompanies the frequency synchronization.
  These results implies that on the small-world network
  the transition of swing oscillators would be different from that of Kuramoto oscillators.

  The hysteretic phenomena have been considered as the triggers of
  the instantaneous power failures$^{5,11,15)}$.
  In the present study the hysteresis of the swing equation 
  seems to disappear provided the network has appropriate
  small-world connections.
  This result implies that the power grid system would be more stable 
  if the transmission network is built up with small-world connections.
  This expectation coincides with the suggestion that
  the small-world connections of the real networks are optimized 
  results of the stability and basin of synchronization$^{16)}$. 
  Since our coarse-grained model is homogeneous coupling and no phase shift term,
  we neglect the inhomogeneous capacities of transmission lines and reactive power.
  Those factors vary the synchronous properties of Kuramoto oscillators$^{17)}$, 
  and therefore more detail investigations are required in future.

\section*{Acknowledgments}
  This research has been received financial supports from 
  the project of Japan Science and Technology agency, JST-CREST.


\begin{thebibliography}{30}
 \bibitem{Corsi_2004_Andersson_2005}
	 S. Corsi and C. Sabelli, IEEE Power Eng. Soci. General Meeting, 1691, (2004);
	 G. Andersson, P. Donalek, R. Farmer, N. Hatziargyriou, I. Kamwa, P. Kundur, N. Martins,
	 J. Paserba, P. Pourbeik, J. Sanchez-Gasca, R. Schulz, A. Stankovic, C. Taylor and
	 V. Vittal: IEEE Trans. Power sys. {\bf 20}[4], 1922 (2005). 

 \bibitem{Pagani_2013}
	 G. A. Pagani and M. Aiello, Physica A {\bf 392}, 2688 (2013) .
	
 \bibitem{Watts_book}
	 D. J. Watts, {\it Small Worlds: The Dynamics of Networks Between Order and Randomness}, 
	 ed. P. W. Anderson, J. M. Epstein, D. K. Foley, S. A. Levin and M. A. Nowak
	 (Princeton Univ. Press., Princeton, 2003) p.3.
	 
 \bibitem{Watts_1998}
	 D. J. Watts and S. H. Strogatz, Nature {\bf 393}, p.440 (1998).	 

 \bibitem{Kundur_book}
	 P. Kundeur, {\it Power System Stability and Control},
	 ed. N. J. Balu and M. G. Lauby,
	 (McGraw-Hill Press., New York, 1994) p.1.

 \bibitem{Filaterlla_2008}
	 G. Filatrella, A. H. Nielsen, and N. F. Pedersen,
	 Eur. Phys. J. B {\bf 61}, 485, (2008).

 \bibitem{Kuramoto_book}
	 Y. Kuramoto, {\it Chemical oscillations, waves and turbulence}, 
	 (Springer-Verlag, New York, 1984) p.1.

 \bibitem{Tanaka_1997a_Tanaka_1997b}
	 H. Tanaka, A. J. Lichtenberg and S. Oishi,
	 Physica D {\bf 100}, 279 (1997);
	 H. Tanaka, A. J. Lichtenberg and S. Oishi, Phys. Rev. Let. {\bf 78}[11], 2104 (1997).

 \bibitem{Dorfler_Bullo_2012}
	 F. D\"orfler and F. Bullo, SIAM J. Control Optim. {\bf 50}[3], 1616 (2012).

 \bibitem{Rohden_2014}
	 M. Rohden, A. Sorge, D. Witthaut and M. Timme, CHAOS {\bf 24}, 013123 (2014).	 

 \bibitem{Rohden_2012_Motter_2013}
	 M. Rohden, A. Sorge, M. Timme and D. Witthaut, Phys. Rev. Lett. {\bf 109}, 064101 (2012);
	 A. E. Motter, S. A. Myers, M. Anghel and T. Nishikawa, Nature Phys. {\bf 9}, 191 (2013).

 \bibitem{Hong_2002}
	 H. Hong, M. Y. Choi and B. J. Kim, Phys. Rev. E {\bf 65}, 026139 (2002).	 

 \bibitem{Um_2014}
	 J. Um, H. Hong and H. Park, Phys. Rev. E {\bf 89}, 012810 (2014).

 \bibitem{Ji_2013}
	 P. Ji, T. K. DM. Peron, P. J. Menck, F. A. Rodrigues and J. Kurths, 
	 Phys. Rev. Lett. {\bf 110}, 218701 (2013).	 
 
 \bibitem{Dobson_Chiang_1989_Susuki_2011}
	 I. Dobson and H.-D. Chiang, Sys. Cont. Lett. {\bf 13}[3], 253-262 (1989);	 
	 Y. Susuki, I. Mezic and T. Hikihara, J. Nonlinear Sci. {\bf 21}, 403-439 (2011).
	 
 \bibitem{Menck_2013}
	 P.J. Menck, J. Heitzig, N. Marwan and J. Kurths, Nature phys. {\bf 9}, 89 (2013).
	 
 \bibitem{Strogatz_1989_Ko_2008} 
	 S. H. Strogatz, C. M. Marcus, R. M. Westervelt and R. E. Mirollo
	 Physica D  {\bf 36}, 23 (1989);	 
	 T.-W. Ko and G. B. Ermentrout,	 Phys. Rev. E {\bf 78}, 016203 (2008).

\end{thebibliography}
\end{document}